\documentclass[12pt]{iopart}

\usepackage{iopams}
\usepackage{graphicx}
\begin{document}

\title{Coherence simplices}
\author{Tapio P. Simula and David M. Paganin}
\address{School of Physics, Monash University, Victoria 3800, Australia}

\begin{abstract}
Coherence simplices are generic topological correlation-function
defects supported by a hierarchy of coherence functions. We
classify coherence simplices based on their topology and discuss
their structure and dynamics, together with their relevance to
several physical systems.\end{abstract} \maketitle

\section{Introduction}

The quantized vortex is an archetypal topological defect which
arises in a variety of physical systems including superfluids,
superconductors, optical speckle fields, coherent fields in the
focal volumes of lenses, coherent optical fields scattered from
sharp edges, eigenmodes of waveguides and angular momentum
eigenstates of the hydrogenic atom
\cite{DonnellyBook,SCvortices,PaganinBook}. The structure and
dynamics of these conventional quantized vortices bear a close
similarity to classical eddies, while enabling some unique
material properties such as crystallization of magnetic flux in
type-II superconductors and quantized circulation in rotating
superfluids \cite{Abrikosov}.

Let us focus on this quantization of circulation,
restricting the discussion to complex scalar fields such as the
wavefunction of Schr\"{o}dinger wave mechanics or the complex
analytic signal associated with a scalar monochromatic
electromagnetic wave.  Quantized vortices in such
fields are very generic, as was emphasized by Dirac in a
visionary paper from the 1930s \cite{Dirac1931}. When studying 
vortices from a topological perspective one
can ignore the details of the particular differential equation
obeyed by a given complex-valued field $\psi$, including whether the said
equation is linear or non-linear, instead relying on the very
general physical property that $\psi$ must be a continuous and 
single-valued function of position and time.  Using
this assumption alone, Dirac was able to show that quantized
vortices will in general exist in complex fields, and that they
are stable with respect to perturbation.  The essential idea,
here, is that while the complex field is single valued, its phase
need not necessarily be single valued; hence the integral of the
phase around a closed loop need not vanish, and may in fact be an
integer multiple $m$ of $2\pi$ radians.  While we explore this
point in further detail in Section 2, the key point to note here
is that non-zero $m$ heralds the presence of a quantized vortex.

As the previous list of examples shows, quantized vortices in
complex fields have long been present in physics.
Examples include Schr\"{o}dinger's angular-momentum eigenstates of
the hydrogenic atom, Wolther's 1950 treatment of vortices in the
context of the Goos--H\"{a}nchen effect, and the 1952 study by
Braunbek and Laukien investigating the exact solution due to
Sommerfeld, for the diffraction of a plane electromagnetic wave
from an infinite half plane \cite{Dennis2009}.  Nye and Berry's
work on dislocations in wave trains \cite{Nye1974} was
pivotal in bringing the study of quantized vortices to the attention
of the scalar optics community.  The study of such optical
vortices has now blossomed into the area of singular optics, with
several key reviews including Berry \cite{BerryLesHouches}, Nye
\cite{NyeBook}, Soskin and Vasnetsov \cite{Soskin2001}, and Dennis
et al. \cite{Dennis2009}.  The nodal lines which thread vortex
cores can form closed loops or extend to infinity
\cite{Dirac1931}; they can terminate at points where the potential
is discontinuous; and they can even form knots
\cite{Faddeev1997a,BerryDennis2001,Leach2010,Dennis2010,Hietarinta2012a}.

Now, partially coherent optical fields or mixed-state wave
functions are often studied via correlation functions such as the
mutual intensity or the cross-spectral density (for partially
coherent optical fields) \cite{ThickWolf, WolfBook} and the
density matrix (for mixed-state wavefunctions)
\cite{Glauber1963a}.  These correlation functions are
very often continuous single-valued complex functions of pairs of
spatial coordinates. If one recalls Dirac's argument for the existence 
of vortices in complex functions, one has the logical possibility of vortex-type
structures in field {\em correlation} functions.  Such vortical
structures, in two-point correlation functions of both classical
and quantum-mechanical complex fields, are known as coherence
vortices.

Accordingly, quantized vorticity has emerged as a topic of
interest in the context of optical and matter field coherence
\cite{Schouten2003,Gbur2003,Palacios2004,Maleev2004,Swartzlander2004,Wang2006a,Wang2006b,GburVisser2006,GburVisserWolf2004,GuGbur2009,
Marasinghe2010a,Marasinghe2011a,Paganin2011a}.  For a recent
review of coherence vortices, see Gbur and Visser
\cite{GburVisser2010}. Here the term `coherence vortex' has been
coined to describe phase singularities or nodal lines in the cross
spectral density matrix, and related coherence functions such as
the spectral degree of coherence \cite{Gbur2003}. Coherence
vortices may undergo topological reactions such as splitting,
fusing and pairwise creation/annihilation \cite{GuGbur2009}.  They
may be spontaneously formed in a Young-type three-pinhole
interferometer \cite{GburVisserWolf2004}, and via Mie scattering
from one or several spheres
\cite{Marasinghe2010a,Marasinghe2011a}. Coherence vortices have
also been studied in the context of linear optical imaging systems
and the focal volume of aberrated lenses \cite{GburVisser2006}.
Recently, it was shown that coherence vortices emerge even in
systems with only one spatial dimension where conventional
vortices are manifestly absent \cite{Paganin2011a}. Coherence
vortices have been observed experimentally in
optical fields \cite{Palacios2004,Swartzlander2004,Wang2006a}.

All of the previously cited work refers to screw-type and/or
edge-type topological defects in two-point coherence functions.
The desire to investigate both higher-order correlation functions
and more complex topological correlation defects has motivated the present
investigation.  Here we develop the notion of a `coherence
simplex' as a generalization of the concept of a coherence vortex,
thereby categorizing topological defects emerging in generic
$p$-point complex correlation functions.

We close this introduction with an overview of the remainder of
the paper.  Section 2 gives a brief background on conventional
quantized vortices as screw-type phase singularities in continuous single-valued complex
functions. In Sec.~3 we define a coherence simplex as a
generalized form of coherence vortex, and in Sec.~4 we categorize
a hierarchy of coherence simplices. Finally, in Sec.~5 we discuss
the structure and dynamics of coherence simplices, followed by
 a discussion and concluding remarks in Sec.~6.

\section{Conventional quantized vortex}
To be self-contained and to facilitate further discussion, we
briefly review some mathematical properties of conventional
quantized vortices. A generic continuous single-valued complex scalar function $f$ of
two real variables $(x,y)$ can, without loss of generality, be cast in
the Madelung form
\begin{equation}\nonumber
f(x,y)=|f(x,y)|e^{i \theta(x,y)},
\end{equation}
\noindent where $\theta(x,y)$ is a real valued phase function
\cite{Madelung1926}. If in the vicinity of some point $(x_0,y_0)$
the phase function has the form
\begin{equation}
   \theta (x-x_0,y-y_0)= m \arctan\left(\frac{y-y_0}{x-x_0}\right) + B(x,y),
\end{equation}
\noindent where $m$ is an integer and $B(x,y)$ is any real
function that is analytic in the vicinity of $(x_0,y_0)$, then the
phase singularity at $(x=x_0,y=y_0)$ is called a point vortex.
Physically, we can view $B(x,y)$ as a continuous background
deformation, upon which sits the screw-type phase dislocation
given by term proportional to $m$.  The amplitude $|f(x_0,y_0)|$
must vanish at the location of the phase singularity to ensure
single valuedness of the complex function. In three-dimensional
systems such zero-amplitude phase singularities are nodal lines or
line vortices which may close on themselves forming loops or
vortex rings
\cite{DonnellyBook,Dirac1931,NyeBook,Feynman1955a,PethickSmith,Holleran2006} or
even vortex knots \cite{Faddeev1997a,BerryDennis2001,Leach2010,Dennis2010,Hietarinta2012a}.

In superfluids, the gradient of the phase function $\nabla\arg[\psi({\bf r})]$ of the complex
order parameter field $\psi({\bf r})$ describing the system can be associated with
the velocity field $v_s({\bf r})$ of the superfluid. Since the curl of the
gradient $\nabla\times\nabla A({\bf r})$ of any vector field $A({\bf r})$ is identically zero unless the field
contains singularities, applying Stokes' law directly leads to the
quantization of circulation of the superflow
\begin{equation}
\oint_{\Omega}\nabla v_s({\bf r}) \cdot {\rm d}{\bf{l}}
   = m\kappa,
\end{equation}
where $m$ is an integer and $\kappa$ is the quantum of
circulation. In helium II and atomic Bose--Einstein
condensates it is the circulation of atoms that is quantized
\cite{DonnellyBook,PethickSmith,PitaevskiiBook}. In superconductors magnetic flux is
quantized and vortices form due to the motion of
Cooper pairs \cite{SCvortices,TsunetoBook,LeggettBook}, and in coherent optical fields
the photons can
form optical vortices \cite{NyeBook,Holleran2006}. It is also
possible to convert optical to matterwave vortices and vice versa
 \cite{Andersen2006a,Simula2008b}.

From a mathematical perspective, any single-valued continuous
complex function of at least two real variables may
possess screw-type phase singularities of this kind
\cite{Dirac1931} and hence vortices should be expected whenever a
model of a physical system involves continuous complex functions.
It is particularly interesting to investigate the vortex
properties of complex $p$-point correlation functions of quantum
fields, to which topic we now turn.

\section{Coherence simplices}

The previous section has restricted consideration to vortices in
field functions, considered as a function of position.  Other
control parameters such as time may be present, but this does not
change the fact that the vortices of Sec. 2 `swirl' in physical
space; in order to swirl in physical space, either two dimensions
$(x,y)$, or three dimensions $(x,y,z)$, are required.  Yet, as
was emphasized in the first two sections of this paper, the
existence of vortices follows directly from one having a
continuous single-valued complex function.  This raises the
logical possibility that complex two-point correlation functions,
which are continuous functions of two spatial coordinates, $(x_1,x_2)$, 
$(x_1,y_1,x_2,y_2)$ or $(x_1,y_1,z_1,x_2,y_2,z_2)$, together with
any other relevant control parameters such as time
coordinates, may also possess vortical structures in the phase of
the said correlation function. However, these vortical
correlation-function phases---termed `coherence vortices'
\cite{Schouten2003,Gbur2003,Palacios2004,Maleev2004,Swartzlander2004,Wang2006a,Wang2006b,GburVisser2006,GburVisserWolf2004,GuGbur2009,GburVisser2010}---will
`swirl' in spaces of high-dimensionality, rather than in physical
space. Moreover, these topological correlation-function phase maps
are not restricted to two-point correlation functions, but may
evidently be generalized to $p$-point correlation functions.  
Such structures, termed coherence
simplices, are the topic of this section.

Let us adopt the language of second quantization
\cite{FetterWalecka} and introduce Heisenberg field operators
$\hat{\psi}^\dagger({\bf r},t)$, $\hat{\psi}({\bf r},t)$ which
respectively create and annihilate an excitation of the field,
such as a particle, at space--time point $({\bf r},t)$.
Furthermore, a bosonic field $\hat{\psi}({\bf r},t)$ obeys the
canonical commutation relations
\begin{eqnarray}
    && [\hat{\psi}({\bf r},t),\hat{\psi}^\dagger({\bf r}',t') ] =
\delta({\bf r}-{\bf r}')\delta(t-t'), \\ \nonumber
  &&[\hat{\psi}({\bf r},t),\hat{\psi}({\bf r}',t')] =
[\hat{\psi}^\dagger({\bf r},t),\hat{\psi}^\dagger({\bf r}',t')
  ]=0,
\end{eqnarray}
\noindent with similar relations holding for fermionic fields with
the commutators replaced by anticommutators \cite{MandlShawBook}.

Consider the $p$-th order field correlation
\cite{Glauber1963a,PitaevskiiBook,FetterWalecka,MandlShawBook}
(cf. many-body density matrix, Glauber quantum coherence
functions, Green functions or Feynman propagators)
\begin{eqnarray}
&&g^{(p)}(x_1,x_2,\ldots x_{2p}) =\langle \hat{\psi}^\dagger({\bf r}_1,t_1)\hat{\psi}^\dagger({\bf r}_2,t_2)\ldots\\
&&\hat{\psi}^\dagger({\bf r}_p,t_p)\hat{\psi}({\bf r}_{p+1},t_{p+1})\hat{\psi}({\bf r}_{p+2},t_{p+2}) \ldots \hat{\psi}({\bf r}_{2p},t_{2p})  \rangle \nonumber
\label{cohfun}
\end{eqnarray}
where $x_i$ refers to both space and time coordinates and the angular brackets denote quantum statistical average. We augment this by further defining
\begin{equation}
\langle \hat{\psi}({\bf r})\rangle=g^{(0)}(x_0)
\end{equation}
\noindent which may acquire finite values through the spontaneous
symmetry breaking mechanism \cite{LeggettBook}.

With these prerequisites, we may define a \emph{ $p$-th order
vortical coherence simplex} as a generalized vortex in the complex
valued correlation function $g^{(p)}(x_1,x_2,\ldots x_{2p})$. For
such vortical coherence simplices, coherence circulation is quantized via
\begin{equation}
\oint_{O}\nabla \arg[g^{(p)}(x_1,x_2,\ldots x_{2p})] \cdot {\rm d}{{\bf l}_{2pD}}
   = m\kappa_p,
\label{cohcirc}
\end{equation}
where ${\rm d}{{\bf l}_{2pD}}$ is a line element along an oriented closed curve $O$ in a $2pD$-dimensional space, where $D$ is the physical dimension of the space, and
the constant $\kappa_p$ is a quantum of coherence circulation. 

As in the case of conventional vortices
\cite{PethickSmith,Freund1999,Simula2002a}, coherence simplices
with winding number $|m|>1$ are likely to be
topologically unstable with respect to perturbations (cf.
\cite{GuGbur2009}), and split into a number of lower
winding-number excitations. By the term coherence
simplex we generically refer to any kind of topological
defect in any correlation function and thus include e.g.
soliton-like structures.

In particular, the $p=0$ VCS reduces to a conventional quantized
vortex in two- or three-dimensional space. It is immediately clear
that if the field $\hat{\psi}({\bf r})$ contains a conventional
vortex, it may also show up as a coherence simplex in the
associated correlation functions (see \cite{Gbur2003,Palacios2004,Swartzlander2004,Wang2006a}). However,
coherence simplices and VCSs in particular, may also appear when
the underlying field itself possesses no conventional vortices. In
fact, VCSs emerge even when $\hat{\psi}({\bf r})$ has only one
spatial dimension and therefore could not exhibit conventional
vorticity even in principle \cite{Paganin2011a}.

In the generalized core, or hypercore, of the VCS $g^{(p)}$ vanishes and the
phase winds an integer $m$ times $2\pi$ around it. While a
conventional material vortex is a zero probability of finding a
particle in some point which defines the location of the vortex
core, the hypercore of a VCS is a zero probability of finding
coherence between two (or more) space--time points. The coherence
simplex can also be described as a topologically unavoidable loss
of coherence between sets of space--time coordinates due to
quantization of circulation of the flow of coherence.
In particular, consider a line integral, (\ref{cohcirc}),
 which yields a nonzero coherence circulation winding number $m$.
 Even if the quantum field maintains full coherence at every point
 along and outside the closed curve $O$, the topology of the problem
dictates there to be at least one generalized point inside $O$ at which the system
is fully incoherent, as is further clarified in Sec. 5.

\begin{figure}
\includegraphics[width=0.9\columnwidth]{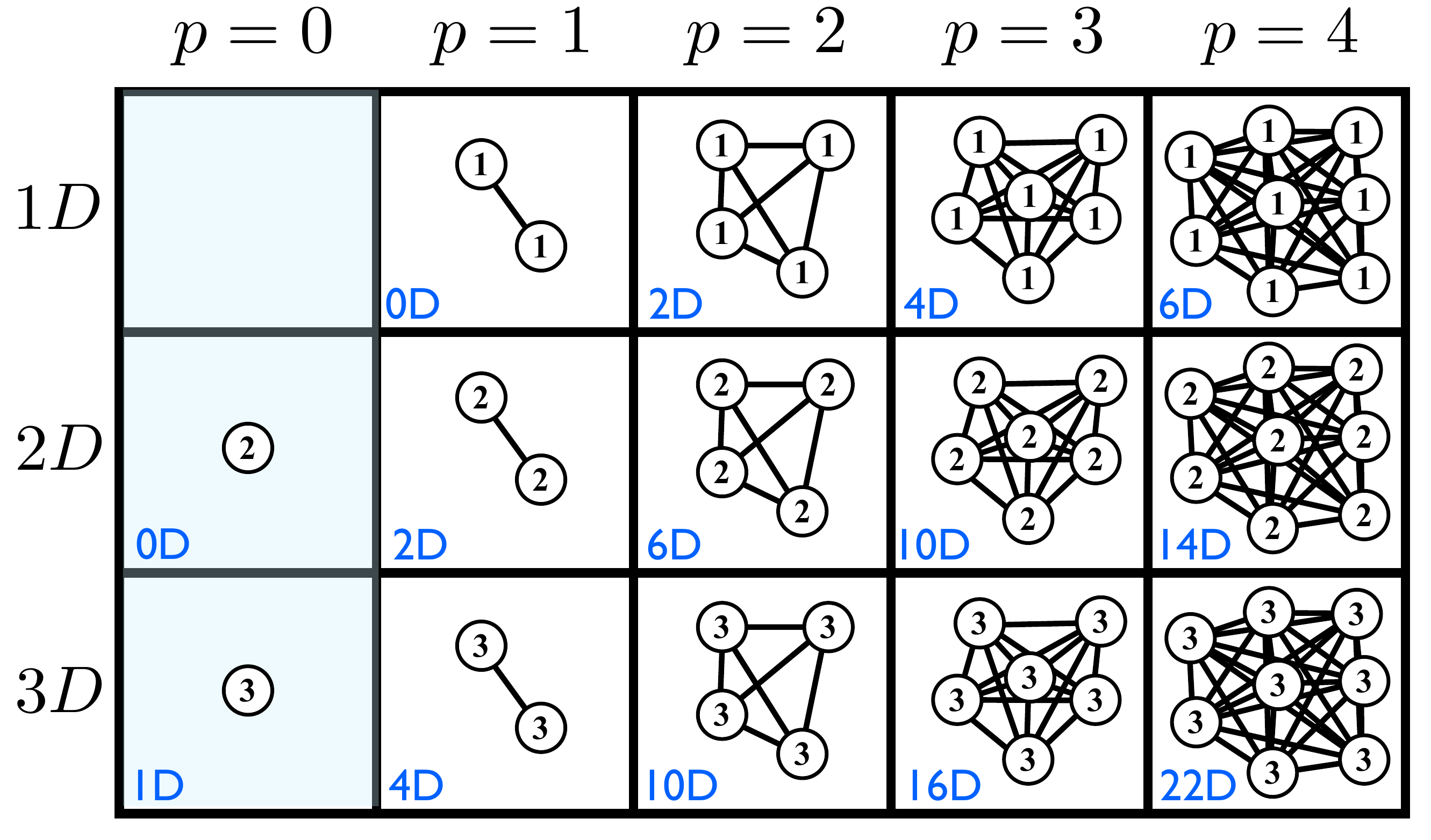}
\caption{Hierarchies of vortical coherence simplices in the space
spanned by spatial dimension versus the order $p$ of the coherence
function. The lower left corner of each matrix element shows the
codimension $C_v$ of the vortical coherence simplex and the diagrams depict the simplex structure of the underlying coherence
function. The numbers inside the bullets denote the physical dimension of the space.}\label{fig1}
\end{figure}

\section{Hierarchies of coherence simplices}

Figure \ref{fig1} illustrates the simplest topologies in the
infinite hierarchy of possible coherence simplices. The
dimensionality of the physical space is depicted on the vertical
axis and the horizontal axis labels the order $p$ of the
correlation function $g^{(p)}(x_1,x_2\ldots x_{2p})$ in which
the coherence simplex is embedded (see e.g. (\ref{cohfun})).
The dimension $M$ of the embedding space of
a VCS is $M=D$ for $p=0$ and $M = 2p\times D$ otherwise, where the
dimension of the physical space is denoted by $D$. A generic
coherence simplex has a codimension
\begin{equation}\nonumber
C = M - N
\end{equation}
\noindent where $N$ is the number of degrees of freedom consumed
by the topological defect type. Vortical coherence simplices
categorized in figure~\ref{fig1} therefore have a codimension $C_v =
M - 2$ which is denoted in the lower left corner of each pane.
The diagram in each pane illustrates the simplex structure of the correlation
function underlying the topological defect. Solitonic coherence
simplices have a codimension $C_s = M - 1$ and, unlike coherence
vortices \cite{Schouten2003,Gbur2003}, exist also in the $D=1,p=0$ case. However, solitons are
not in general topologically protected in the way vortices are and
therefore their stability with respect to perturbations \cite{NyeBook} is determined by energetic
considerations rather than their topology. Higher order $N>2$
coherence simplices can be categorized using similar diagrams to
that shown in figure~\ref{fig1}.

The first (shaded) column in figure \ref{fig1} corresponds to the
conventional vortices, which are manifestly absent in systems with
one spatial dimension. In two-dimensional systems these
conventional vortices are point charges whereas in the
three-dimensional case they correspond to line-like vortex filaments. The
second column in figure \ref{fig1} contains all possible VCSs
supported by the two-point correlation functions. In particular,
in 1D these correspond to coherence vortices \cite{Paganin2011a}
and in 2D they are sheet-like structures.

Coherence simplices can be constrained to lower dimensional spaces
revealing a correspondingly simpler topological structure. For
example, slicing through a conventional 3D vortex line results in
a 2D point vortex. Similarly, fixing three of the six coordinates
$(x,y,z,x',y',z')$ of a first order $p=1$ VCS in a physical space with
three spatial dimensions $D=3$ reveals the VCSs as a one-dimensional
nodal line object, whereas if all of the coordinates are left free
the VCS is a four dimensional object. Most of the coherence
simplices discussed in the literature so far
\cite{Schouten2003,Gbur2003,Wang2006a,Wang2006b,Marasinghe2010a,Marasinghe2011a}
correspond to such reduced-dimensional or constrained VCSs.
Curiously, as is evident in figure \ref{fig1}, a line-like $C_v=1$ VCS only
exists for $p>0$ in the presence of constraints fixing one or more degrees of
freedom but never as a free object. On the other hand, under
suitable constraints any VCS with finite codimension $C_v>0$ can
be constrained to appear as a line-like coherence vortex.

\section{Structure and dynamics of coherence simplices}

The zeroth-order simple vortical coherence simplices are simply
conventional vortices with vanishing field amplitude in the
vicinity of a screw-type phase singularity. In other words they correspond
to a hole around which the entities described by the field, such
as atoms or photons, circulate. In contrast, the higher order
($p>0$) VCSs typically have finite probability of finding corpuscles inside
their hypercores, despite the fact that the corresponding
coherence function vanishes there. As in the case of conventional
vortices, due to energetic considerations multi-quantum ($|m|>1$)
VCSs are likely to split into multiple lower order VCSs and
multiple VCSs in equilibrium may arrange in a VCS lattice whose
structure may be that of an Abrikosov-type triangle shape, square,
etc. depending on the details of the defect topology and
interactions.

\begin{figure}
\includegraphics[width=0.9\columnwidth]{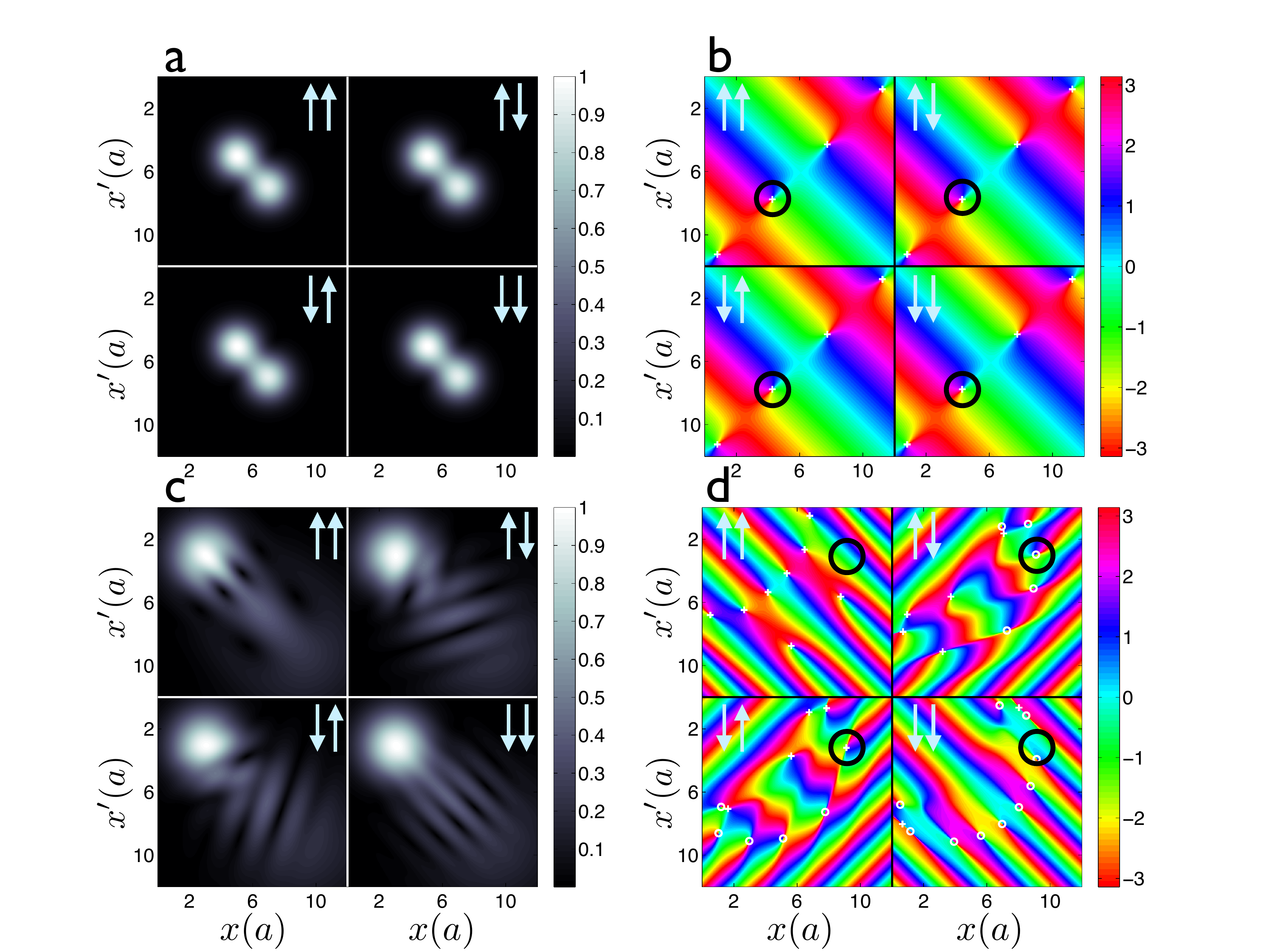}
\caption{Vector coherence simplices in a two-member (a,b) and three-member (c,d) ensembles of two-state wavefunctions defined in the text. The four different spin-spin correlations are denoted in the upper corners of each frame in each subfigure. Subfigures (a) and (c) show the coherence density and (b) and (d) show the phase of the coherence function. Two different kinds of vector coherence defects are circled in frames (b) and (c). The constant $a$ is an arbitrary spatial length scale.}\label{fig2}
\end{figure}

If the complex field supporting the VCSs has multiple internal
degrees of freedom, the field operator becomes a multicomponent
spinor
$\hat{\psi}_\sigma^\dagger({\bf r},t)=(\hat{\psi}_1^\dagger({\bf r},t),\hat{\psi}_2^\dagger({\bf r},t),\ldots \hat{\psi}_\sigma^\dagger({\bf r},t))^T$, where $T$ denotes transpose and $\sigma$ is a spin index, and hence the coherence functions
have a tensor structure $g_{\sigma\sigma'}^{(p)}(x_1,x_2\ldots x_{2p})$.
Through this construction such coherence functions will in general admit
coherence singularities analogous to monopoles, skyrmions, fractional
vortices, hedgehogs, sheets, textures, knots etc. \cite{VolovikBook,
Huhtamaki2009a,UedaBook,Simula2011a}. In such spinor fields with
vector order parameter, continuous coherence simplices for which
even the coherence density does not have to vanish in the spinor VCS
hypercores are possible.

Figure 2 is an illustration of vector coherence defects produced by the mixture of 
(a,b) two and (c,d) three traveling two-component Gaussian wave packets of the form
$\psi_n =A_n e^{-((x-x_n) / \sqrt{2}\mu_n)^2}( e^{ik^\uparrow_n x},e^{ik^\downarrow_n x})^T/\mu_n$, where $A_n$ is a normalization constant. To produce figure 2 (a,b) we used dimensionless values $x_1=-1,k^\uparrow_1=k^\downarrow_1=0.5,\mu_1=1$ and $x_2=1,k^\uparrow_2=k^\downarrow_2=-0.4,\mu_2=1$ and for (c,d) $x_1=3.0, k^\uparrow_1= -2.0, k^\downarrow_1=2.0, \mu_1=3.0$,  
$x_2=0, k^\uparrow_2=  k^\downarrow_2=-1.0, \mu_2=2.0$ and $x_3=-3.0, k^\uparrow_3= 1.5, k^\downarrow_3=-1.5, \mu_2=1.5$.

Frames (a) and (c) show the resulting coherence densities $|g_{\sigma\sigma'}^{(1)}(x,x')| $ and (b) and (d) are
the corresponding phase functions. The circles in (b) and (c), respectively
mark the location of a winding number $w=\{1,1,1,1\}$ and
$w=\{0,-1,1,0\}$ defects where the generalized winding number 
$w=\{m_{11},m_{12},\ldots m_{\sigma\sigma'}\}$ is expressed as vector
of winding numbers corresponding to the different spin-spin correlation functions.

Coherence simplices are generically dynamical objects flowing with
the coherence current \cite{Wang2006a}. When perturbed,
equilibrium coherence simplices are expected to reveal elementary
excitation modes such as helical Kelvin waves
\cite{PitaevskiiBook,Lordi,Simula2008a} and Tkachenko shear waves
\cite{Tkachenko,Sonin,Coddington,Simula2010a} and their VCS
displacement-wave generalizations.  In this context, figure~3 of
Marasinghe {\em et al.} \cite{Marasinghe2011a} displays a clear
signature of oscillatory dynamics, similar to the Crow instability
of antiparallel conventional vortices \cite{Crow,Simula2011b},
between coherence vortex--antivortex pairs. Non-equilibrium
coherence simplex dynamics may reveal VCS reactions such as VCS
intercommutation or reconnection events and coherence-simplex
turbulence (cf. figure 4).

In contrast to the topological arguments given earlier, a study of
the {\em dynamics} of coherence simplices presupposes knowledge of
the equations of motion underlying a given field
$\hat{\psi}({\bf r},t)$. Suppose that this equation, for the field
supporting a given coherence simplex or simplices, is given by the
Heisenberg equation of motion \cite{FetterWalecka}
\begin{equation}
i\hbar\frac{\partial}{\partial t}\hat{\psi}({\bf r},t)=[\hat{\psi}({\bf r},t),\hat{H}],
\label{leq}
\end{equation}
where $\hat{H}$ is the Hamiltonian operator. From here on we focus
on a free-field case (interactions alter the dynamics of coherence
simplices but do not invalidate the topological considerations)
 $\hat{H}=-\hbar^2\nabla^2 /2m$ and define a
linear differential operator
\begin{equation}
\mathcal{L}=i\hbar\frac{\partial}{\partial t}+\frac{\hbar^2\nabla^2}{2m},
\end{equation}
in terms of which (\ref{leq}) may be expressed as
\begin{equation}
\mathcal{L}\hat{\psi}({\bf r},t)=0.
\end{equation}
Acting with the components $k$ of $\mathcal{L}$ on the coherence
function $g^{(p)}(x_1,x_2\ldots x_{2p})$, we obtain a set of
Wolf-like equations \cite{ThickWolf,WolfBook,Wolf1954}
\begin{equation}
\mathcal{L}_k g^{(p)}(x_1,x_2\ldots x_{2p})= 0.
\label{coheq}
\end{equation}
These equations can be cast in a `hydrodynamic' form \cite{Madelung1926} by separating the
real and imaginary parts which yields two sets of coupled
equations
\begin{eqnarray}
{\rm Re}[g^{(p)*}\mathcal{L}_kg^{(p)}]=0,\\
{\rm Im}[g^{(p)*}\mathcal{L}_kg^{(p)}]=0. \nonumber
\label{eq95}
\end{eqnarray}
These correspond to an Euler-like equation and a continuity
equation for coherence flux, which govern the evolution of
coherence in the system and hence the dynamics of the coherence
simplices which move with the coherence flow \cite{Wang2006b}.
The spectrum of coherence excitations or coherence quasi-particles can be
found by linearising these equations \cite{PethickSmith,PitaevskiiBook}.

The formal solution to (\ref{coheq}) is
\begin{eqnarray}
&&g^{(p)}(x_1,x_2\ldots x_{2p}) = \int \int\ldots\int dx_1'dx_2'\ldots dx_{2p}'\\
&\times&G(x_1,x_2\ldots x_{2p},x_1',x_2'\ldots
x_{2p}')g^{(p)}(x_1',x_2'\ldots x_{2p}'), \nonumber
\label{propagate_coherence}
\end{eqnarray}
which evolves the coherence function into the future using the
forward-time many-body propagator
\begin{eqnarray}
&&G(x_1,x_2\ldots x_{2p},x_1',x_2'\ldots x_{2p}') =\\
&&-i\langle [\hat{\psi}(x_1)\ldots\hat{\psi}(x_{2p})
\hat{\psi}^\dagger(x'_{1})\ldots \hat{\psi}^\dagger(x'_{2p})]
\rangle. \nonumber\label{many_body_propagator}
\end{eqnarray}
Here we assume a time-ordering in which all primed time coordinates lie in the future of their corresponding unprimed coordinates. Equation~(\ref{propagate_coherence}) may be viewed as a generalized
Huygens-type formula for the propagation of coherence \cite{ThickWolf}, the
propagation occurring from one set of equal-time space--time
points $(x_1',x_2',\ldots,x'_{2p})$, to a different set of
equal-time future space--time points $(x_1,x_2,\ldots,x_{2p})$.
For the special case of two-point correlation functions in one
spatial dimension, this may be visualized as shown in
figure~3(a).  This depicts the propagation of the two-point
equal-time one-spatial-dimensional correlation function $g^{(1)}$,
from its boundary value over the hyperplane $t_1=t_2=t$, to its
boundary value over the hyperplane $t_1=t_2=t+\tau$, $\tau\ge 0$.
Spatial coordinates are denoted with uppercase $X$, to distinguish
from space--time coordinates denoted by a lowercase $x$.  In this
low-dimensional example, the coherence function over an
infinitesimal patch at $A=(\tilde{X}'_1,t_1=t,\tilde{X}'_2,t_2=t)\equiv
(x'_1,x'_2)$ with infinitesimal area $d\tilde{X}'_1 d\tilde{X}'_2=dx'_1 dx'_2$ in
the $X'_1-X'_2$ plane (denoted $\Pi_1$), propagates to give the
contribution $dx'_1 dx'_2 G(x_1,x_2,x'_1,x'_2)g^{(1)}(x'_1,x'_2)$
to the coherence over the entire $X_1-X_2$ plane (denoted
$\Pi_2$).  This contribution is the coherence-theory analogue of the
concept of a Huygens elementary wavelet, as embodied in the
Huygens--Fresnel principle \cite{ThickWolf}. Every such contribution (from the
points $A,B,C$ etc. in figure~3(a)) is then summed, via
Eq.(\ref{propagate_coherence}), to give the propagated coherence
function $g^{(1)}(x_1,x_2)$ over the plane $\Pi_2$.

\begin{figure}
\includegraphics[width=0.9\columnwidth]{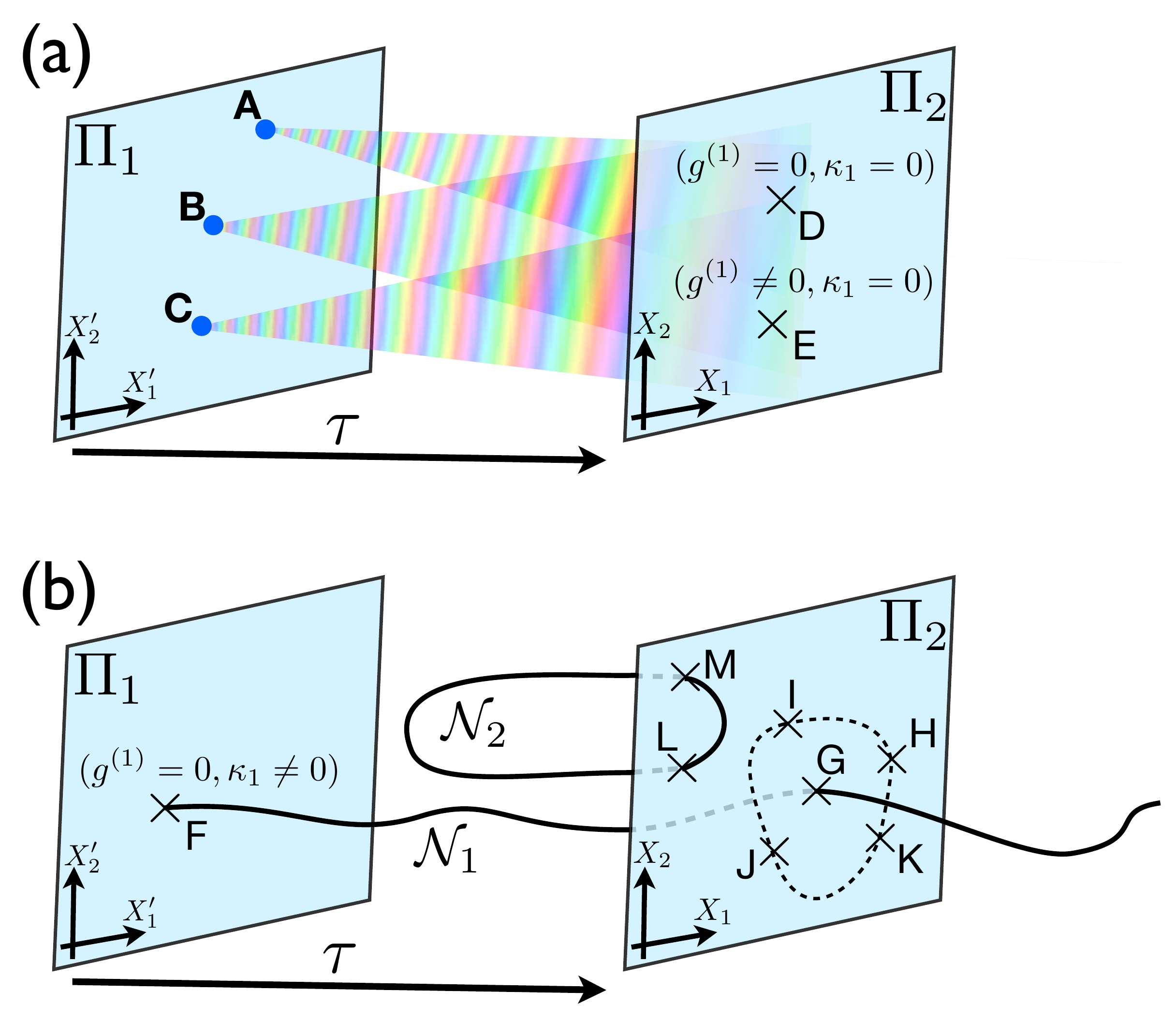}
\caption{Coherence simplices illustrated.  As discussed in the
main text, panel (a) shows a Huygens-type construction for the
propagation of two-point correlation functions from hyperplane
$\Pi_1$ to hyperplane $\Pi_2$.  Three Huygens wavelets are shown,
emanating from points $A$, $B$ and $C$; $\tau$ is a control
parameter, which can be pictured as a generalized propagation
distance, whose increasing value continuously evolves $\Pi_1$ into $\Pi_2$.
Panel (b) shows an open nodal line $\mathcal{N}_1$ which threads
coherence vortices located at points $F$ and $G$, with
$\mathcal{N}_2$ denoting a closed coherence-vortex loop.  A third
possibility, that of a nodal-line knot $\mathcal{N}_3$  or a coherence Hopfion (cf.
\cite{Faddeev1997a,BerryDennis2001,Leach2010,Dennis2010,Hietarinta2012a}), is not shown.}
\label{fig2}
\end{figure}

This superposition of elementary Huygens-type wavelets may lead to
points $D$ at which $g^{(1)}$ vanishes, due to what may be termed
`complete destructive interference of the coherence-function
wavelets' at such a point.  Coherence vortices ($\mathcal{N}_1$ and
$\mathcal{N}_2$ in figure~\ref{fig2}(b), which appear as
points $F,G,L$ and $M$ in the planes $\Pi_1$ and $\Pi_2$), for which
a zero of $g^{(1)}$ is accompanied with a non-zero value of $\kappa_1$,
are a special case of such complete destructive interference of coherence.
More `regular' points such as $E$ in figure~3(a) (and $H,I,J$ and $K$ in figure~\ref{fig2}(b))
correspond to non-zero values for $g^{(1)}$.

By encircling the coherence vortex $G$ in figure~\ref{fig2}(b), the associated simple smooth closed path
sampled at points $H,I,J$ and $K$ picks up a phase factor which identifies a vortical
coherence field. If such a path integral yields non-zero $\kappa_1$, even if the field is fully or partially coherent
at points $H,I,J$ and $K$, there must be at least one point such as $G$ in the space enclosed
by the path, where coherence vanishes. A low-dimensional example, of this, is studied in
Ref. \cite{Paganin2011a}. Generically, the hypercore of a coherence simplex corresponds to a zero in the
function $g^{(p)}(x_1,x_2\ldots x_{2p})$ with an associated non-zero coherence
circulation $\kappa_p$. Note that the propagator,
(\ref{many_body_propagator}),  being a correlation function itself,
may also be vortical.

Figure~3(b) also illustrates the conservation of the
coherence circulation. Consider a coherence vortex $\mathcal{N}_1$
which at time $t$ pierces the $X'_1-X'_2$ plane and is the only
coherence vortex present at that time. As for conventional
vortical systems \cite{DonnellyBook,LeggettBook}, the total
circulation is a topological invariant and therefore must be
preserved during the propagation of the coherence over any time
interval $\tau$. Note, however, that coherence vortex--antivortex
pairs $(L,M)$ may nucleate and/or annihilate in the course of
propagation without affecting the value of the total coherence
circulation. Such coherence vortex pairs appear as a manifestation
of coherence vortex space--time loops, such as $\mathcal{N}_2$,
along which the coherence vanishes.  Coherence-vortex nodal-line
knots (not shown) are also possible; such knotted vortices
have been studied both theoretically and experimentally, for the
case of coherent fields
\cite{Faddeev1997a,BerryDennis2001,Leach2010,Dennis2010,Hietarinta2012a}.

Finally, we demonstrate the evolution of vector coherence simplices, shown in figure 2 (a,b) and confined in a toroid of length $12 a$, 
by propagating the coherence function forward in time as described by (\ref{propagate_coherence}). The nodal lines of
coherence with $m=\pm 1$ are denoted by green and black lines. Coherence vortex waves and loops
of the kind $\mathcal{N}_2$ (the fundamental elements of quantum coherence turbulence) are clearly visible in the figure. Space-time points where green and black lines join correspond to coherence vortex pair creation and annihilation events.

\begin{figure}
\includegraphics[width=0.5\columnwidth]{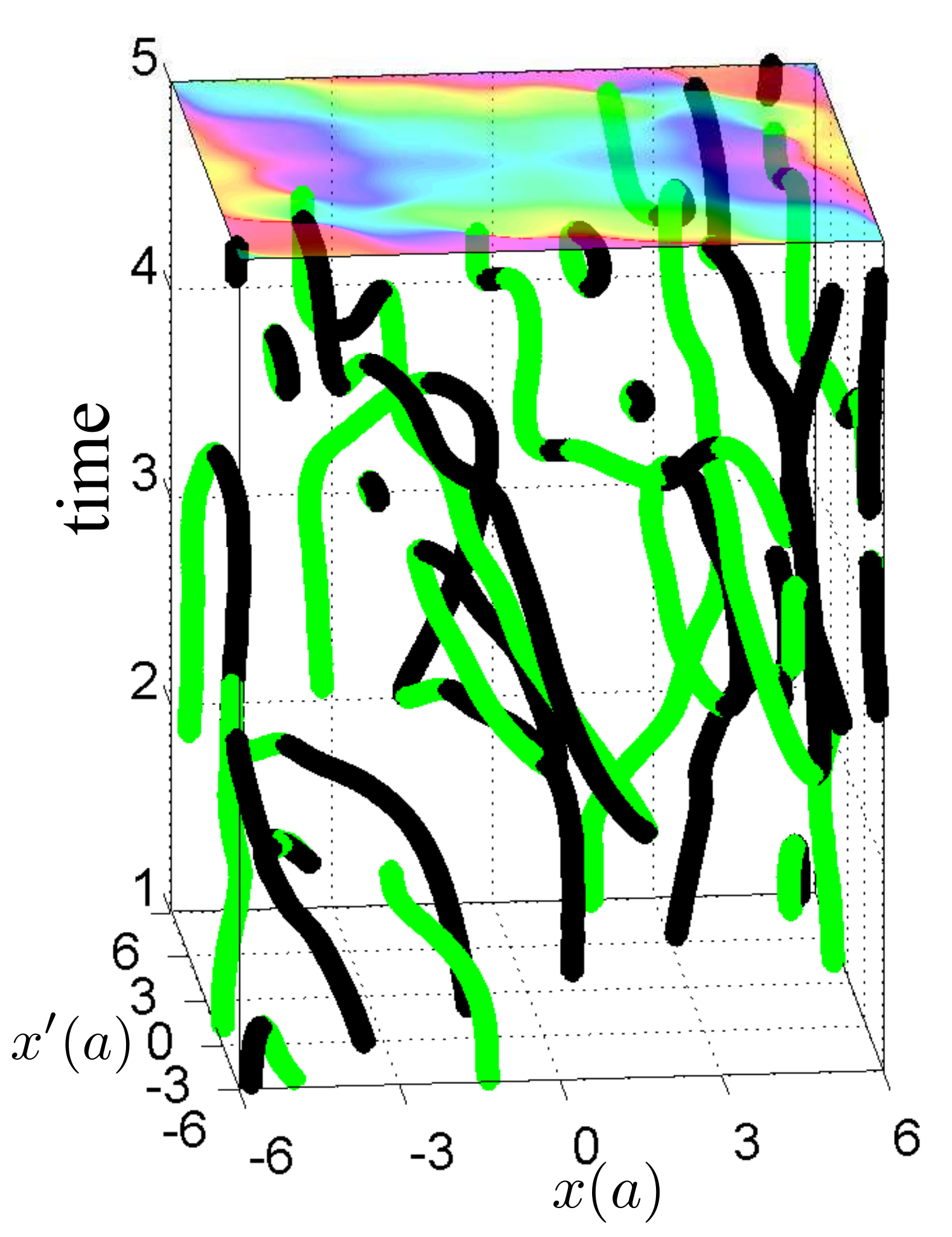}
\caption{Vector coherence dynamics. The nodal lines corresponding to the propagation of the coherence field shown in figure 2a. Due to the symmetry of the problem only the $(\uparrow,\uparrow)$ spin component of the coherence function is shown. Green and black lines correspond to $m=1$ and $m=-1$ coherence circulations, respectively. Coherence vortex loops and coherence vortex displacement waves are clearly visible in the figure. Coherence phase function at time $t=5$ is also plotted (coloring corresponds to that in the figure 2 (b) and (d)).
}\label{fig2}
\end{figure}

\section{Discussion}

We have introduced a hierarchy of coherence simplices. A zeroth-order coherence simplex is nothing but a conventional quantized
vortex in complex wavefunctions describing e.g. coherent optical
fields or superfluid matter. Higher-order vortical coherence
simplices emerge as quantized vortices in generic multi-dimensional correlation
functions. We have discussed the topological structure and
dynamics of such coherence simplices. A hypercore of a coherence
simplex corresponds to a topologically inevitable total
destructive interference of the coherence-function
wavelets. The existence of first-order vortical coherence simplices have already
been verified experimentally using interferometric measurements of
light \cite{Wang2006a}. To our knowledge, they are yet to be
discovered in the quasi-particle and matter-wave counterparts of
coherent light.

Furthermore, the Berezinskii--Kosterlitz-Thouless mechanism, which
enables quasi-long-range order and superfluidity in
two-dimensional systems, relies upon the coherence of
vortex--antivortex pairs \cite{Berezinskii,Kosterlitz}. However,
direct observation of such spontaneously forming vortex dipoles
has remained elusive
\cite{Hadzibabic,Simula2006a,Clade,Tung,Hung}. It would be
interesting to apply the theory of coherence simplices
to such systems and to investigate the role of VCSs in the
superfluid--normal phase transition in low-dimensional systems.
The recently created photon Bose--Einstein condensates
\cite{Klaers2010b} provide a natural platform to study
coherence simplices and their relation to conventional quantized
vortices which are a hallmark of superfluidity and are responsible
for many of their special properties. It is therefore fascinating
to contemplate the analogous idea that higher-order coherence
simplices might underlie and signify some fundamental yet so far
undiscovered physical properties of matter. Finally, it was pointed
out in \cite{Paganin2011a} that coherence simplices
may be associated with the process of quantum mechanical
decoherence---a suggestion which warrants further research.


\end{document}